\documentstyle[aps,prl,epsf,floats,twocolumn,physica]{revtex}
\begin{document}
\draft
\twocolumn[\hsize\textwidth\columnwidth\hsize\csname @twocolumnfalse\endcsname

\vspace{1in}
\begin{center}
\today \\
\vspace{0.5cm}

{\large \bf Metal-Insulator Transition in a Low-Mobility Two-Dimensional 
Electron System} \\
\bigskip
Dragana Popovi\'{c}$^{1}$, A. B. Fowler$^{2}$, and S. Washburn$^{3}$\\
\vspace{0.2cm}
$^{1}${\it National High Magnetic Field Laboratory, 
Florida State University, Tallahassee, FL 32306, USA} \\ 
$^{2}${\it IBM Research Division, T. J. Watson Research Center, Yorktown
Heights, NY 10598, USA} \\ $^{3}${\it Dept. of Physics and 
Astronomy, The University of North Carolina at Chapel Hill, Chapel Hill, NC 
27599, USA }
\end{center}
\hrule
\vspace{0.2cm}
\centerline{\bf Abstract}
\vspace{0.3cm}

We have varied the disorder in a two-dimensional electron system in silicon 
by applying substrate bias.  When the disorder becomes sufficiently low, we
observe the emergence of the metallic phase, and find evidence for a 
metal-insulator transition (MIT): the single-parameter scaling of conductivity
with temperature near a critical electron density.  We obtain the scaling
function $\beta$, which determines the length (or temperature)
dependence of the conductance.  $\beta$ is smooth and monotonic, and linear
in the logarithm of the conductance near the MIT, in agreement with the 
scaling theory for interacting systems.

\vspace{0.5cm}

\vspace{0.5cm}
\hrule
\vspace{1cm}

]

\noindent {\bf 1. Introduction}
\vspace{0.5cm}

Recent experiments~\cite{Krav,dpmit} on a two-dimensional electron 
system (2DES) in Si metal-oxide-\linebreak semiconductor field-effect 
transistors (MOSFETs) have
provided solid evidence for the existence of a metal-insulator transition (MIT)
at zero magnetic field in this 2D structure, raising speculation that this 
transition is driven by electron-electron interactions.  However, the disorder 
in Si MOSFETs at the transition is also strong, as indicated by the 
reported~\cite{Krav,dpmit} values of the critical conductivity 
$\sigma_{c}\sim e^2/2h$ (corresponding to $k_{F}l${\scriptsize 
$\stackrel{\textstyle _<}{_\sim}$}1, where $k_F$--Fermi wavevector, $l$--mean 
free path).  Experiments on mesoscopic Si MOSFETs~\cite{dpmes} and
a recent observation of a MIT in clean ($k_{F}l\approx$40) Si/SiGe
devices~\cite{khalid} with a similar value of the critical electron density
$n_c$, support strongly the view that electron-electron interaction is the
most relevant energy scale at the MIT.  In this paper, however, we focus on 
the role of disorder, and show that the metallic phase emerges only when the 
disorder is sufficiently low.  We also discuss the form of the conductance
scaling function (``beta 
function''~\cite{gang}), which defines the (length, temperature) scale 
dependence of the conductance.  Our results seem to be consistent with the 
suggestion~\cite{newgang} that the 2D metal is not a Fermi liquid.

\vspace{1.0cm}
\noindent {\bf 2. Experiment}
\vspace{0.5cm}

\begin{figure}[t]
\epsfxsize=3.0in \epsfbox{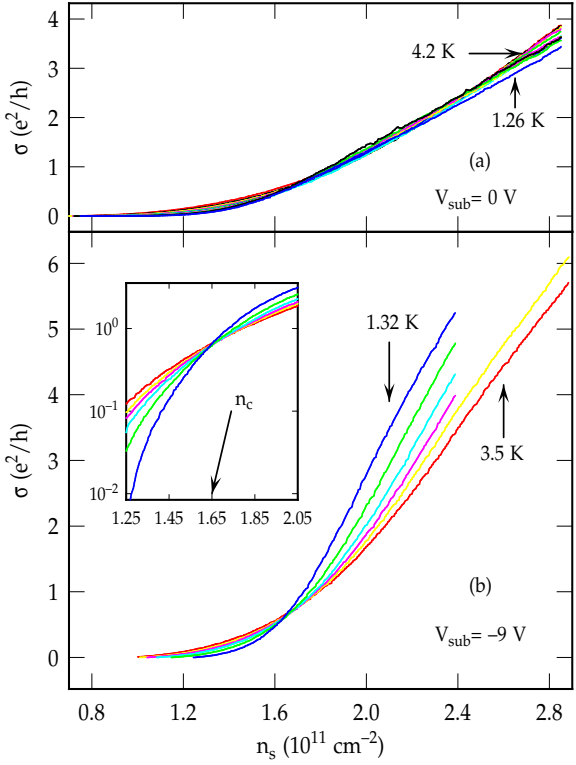}\vspace{5pt}
\caption{Conductivity $\sigma$ of sample \#1 as a function of $n_{s}$ for (a)
$T=$4.2, 3.6, 3.2, 2.8, 2.5, 2.3, 2.09, 1.93, 1.79, 1.67, 1.57, 1.21~K and 
$V_{sub}=0$~V, and (b) $T=$3.5, 3.0, 2.7, 2.3, 1.85, 1.32~K and
$V_{sub}=-9$~V.  In (a), $\sigma$ decreases for all $n_{s}$ as $T$ is lowered.
In (b), $\sigma$ increases as $T$ goes down for all $n_{s}>n_{c}$.  The inset 
shows the same data around $n_{c}$ with $\sigma$ on a logarithmic scale.
\label{dhva1cond}}
\end{figure}

We present results obtained on n-channel MOSFETs fabricated on the (100)
surface of Si doped at $\approx 8.3\times 10^{14}$ acceptors/cm$^{3}$ with 
435~\AA\, gate oxide thickness and oxide charge $\approx 3\times 
10^{10}$cm$^{-2}$.  The samples had a Corbino (circular) geometry with the 
channel length $L=0.4$~mm and width $W=8$~mm.  Conductance $G$ was measured as
a function of gate voltage $V_{g}$ (proportional to carrier density $n_{s}$) 
at temperatures $1.2 < T < 4.2$~K, using a lock-in at a frequency of 
$\sim 100$~Hz and an excitation voltage of 0.3~mV.  The peak mobility, which is
a rough measure of the amount of disorder~\cite{AFS}, was 0.5--0.8~m$^{2}/$Vs 
at 4.2~K.  In these samples, all electronic states are localized, 
in agreement with early studies~\cite{weakloc}.  Fig.~\ref{dhva1cond}(a) shows 
the conductivity $\sigma =GL/W$ for one of the samples (sample \#1), as a 
function of electron density $n_{s}$ for several temperatures.  As $T$ is 
lowered, $\sigma$ decreases for both low and high $n_{s}$, indicating 
insulating behavior.  The temperature dependence is weaker at intermediate
values of $n_{s}$, which is probably a precursor of the critical behavior
discussed below.

By applying the reverse substrate bias we have been able to decrease 
the disorder scattering and increase the mobility at all carrier 
densities~\cite{Alan}, with the peak mobility going up to 
$\approx$1~m$^{2}/$Vs at 4.2~K.
At that point, the mobility becomes sufficiently
high at carrier densities low enough to allow the Coulomb interaction energy 
$U$ to be much greater than the Fermi energy $E_F$~\cite{Krav,newgang}
(in our samples, $U\approx$150~K~$\gg E_{F}\approx$12~K), and we observe the 
metal-insulator transition, as shown in Fig.~\ref{dhva1cond}(b) for 
$V_{sub}=-9$~V: the temperature dependent behavior of $\sigma$ is now reversed 
for all $n_{s}$ above $n_{c}=(1.65\pm 0.02)\times 10^{11}$cm$^{-2}$, as 
expected for metallic behavior.  For $n_{s}< n_{c}$, the 2DES exhibits 
insulating behavior as before.   

\vspace{1.0cm}
\noindent {\bf 3. Scaling analysis}
\vspace{0.5cm}

\begin{figure}[t]
\epsfxsize=3.0in \epsfbox{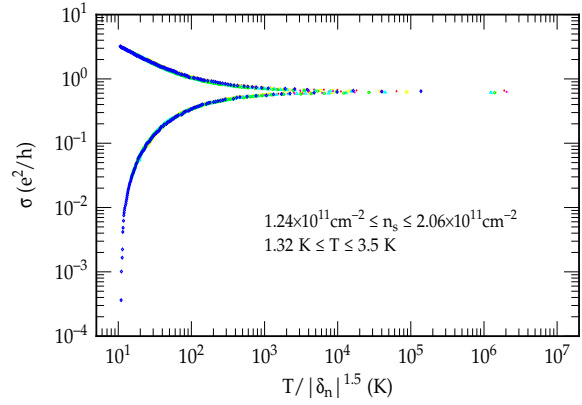}\vspace{5pt}
\caption{Scaling of conductivity with temperature for sample \#1, using the
data shown in Fig.~1(b) in the $n_{s}$ and $T$ ranges given on
the plot.  
\label{bzeroscaling1}}
\end{figure}

Fig.~\ref{bzeroscaling1} shows the scaling of $\sigma$ with temperature near
the transition according to
\begin{equation}
\label{eq1}
\sigma (T,n_{s})=f(|\delta_{n}|/T^{1/z\nu})=\sigma (T/T_{0}),
\end{equation}
with a single parameter $T_{0}$ that is the same function of 
$\delta _{n}\equiv (n_{s}-n_{c})/n_{c}$ on both the metallic and the
insulating side of the transition, $T_{0}\propto |\delta _{n}|^{z\nu}$.
Here $z$ is the dynamical exponent, and $\nu$ is the correlation length
exponent.  
The best collapse is found for $z\nu =1.5\pm 0.1$, in excellent agreement with
other studies on Si MOSFETs~\cite{Krav}.  

A quantity of special interest is the scaling function~\cite{gang}
$\beta (g)=d[\log (g)]/d[\log (L)]$ (where $g=\sigma /(e^2/h)$ is a 
dimensionless conductance), 
which describes the length scale dependence of the 
conductance.  At non-zero temperatures, the effective sample size is 
$L_{eff}\sim T^{-1/z}$~\cite{newgang,LR,Girvin,zcomment}, so that $\beta (g)$ 
determines the
temperature dependence of the conductance.  In particular, the shape of $\beta$
at large $g$ determines the temperature dependence of the metallic phase.
For non-interacting systems and for Fermi liquid systems, $\beta (g)=(d-2)$ for
$g\rightarrow +\infty$ ($d$--dimension), which is equivalent to Ohm's law.
%Obviously, it follows that $\beta (g=+\infty )=0$ in 2D.  
However, it has been 
argued~\cite{newgang} that the 2D metallic phase is {\it not} a Fermi liquid
but some more exotic state.  In general, one does not expect Ohm's law to hold
for non-Fermi liquid metals, as is well known, for example, in the case of a
Luttinger liquid~\cite{GS}.  Instead, conductance could display a non-trivial
length dependence even in the limit $g\rightarrow +\infty$.  Therefore, it is
interesting to examine experimentally the behavior of $\beta (g)$,
in order to obtain clues about the nature of the metallic phase in 2D.
More specifically, if experiments suggest that $\beta\neq 0$ as $g\rightarrow
+\infty$, such behavior would be consistent with a non-Fermi liquid metallic
state.

\begin{figure}
\epsfxsize=3.0in \epsfbox{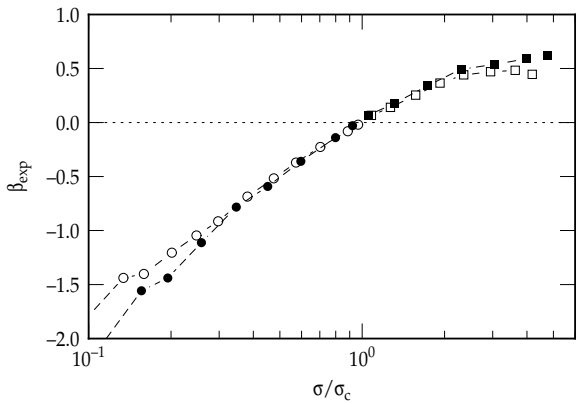}\vspace{5pt}
\caption{$\beta (g)$ vs. conductance on a logarithmic scale.  $\beta_{exp}$ 
was derived from the data such as those shown in Fig.~2: dots were obtained 
from the insulating branch, and squares from the metallic branch.  Solid
symbols: sample \#1, $\sigma_{c}=0.65~e^2/h$; open symbols: sample \#5,
$\sigma_c =0.5~e^2/h$.
\label{betafctn}}
\end{figure}

Fig.~\ref{betafctn} shows an analogous scaling function $\beta_{exp}
 =-d[\log g ]/d[\log T]$ for two of our samples.
$\beta_{exp} (g)$ changes sign for $\sigma =\sigma_c$, as 
expected at the metal-insulator transition.  The shape of $\beta_{exp}$ near
the transition is consistent with the suggestion~\cite{newgang} that, on a 
logarithmic scale, $\beta (g)$ is linear over an appreciable range of
conductances.  $\beta_{exp}$ increases monotonically and, most interestingly, 
may saturate at a value of 0.5 for large $g$.  If this behavior would
persist for arbitrary large values of $g$, it would indicate that the metallic
phase is a non-Fermi liquid, characterized by conductance diverging as
$T^{-1/2}$ when $T\rightarrow 0$.  Another possibility is that there is a 
broad maximum in $\beta (g)$ and that it will turn back down toward the value
of zero in accordance with Ohm's law.  In either case, however, we have a
clear violation of the scaling picture for non-interacting systems~\cite{gang},
where $\beta < 0$ for $d=2$.

The data obtained on high-mobility Si MOSFETs~\cite{Kravscaling} yielded 
$\beta (g)$ with only a monotonic increase with $\log g$, up to a value of 
about 1, in exactly the same range of $g$ as our data (Fig.~\ref{betafctn}).
At higher values of $g$, $\beta (g)$ turned back down toward zero, with no
indication of a saturation.  In that range of $g$, however, the scaling failed
and $\beta (g)$ could not be extracted reliably from the 
data~\cite{Kravscaling,Kravprivate}.  Clearly, it would be 
interesting to study the beta function at much higher values of $g$, for which
systems other than Si MOSFETs might be more appropriate~\cite{khalid}.

\vspace{1.0cm}
\noindent {\bf 4. Conclusion}
\vspace{0.5cm}

In summary, we have shown how the metallic phase can be created by reducing
the disorder scattering on a single sample.  The resulting scaling behavior of
conductivity with temperature provides convincing evidence for the existence of
a metal-insulator transition in this 2D system.  In addition, the results of 
our scaling analysis seem to be consistent with the suggestion that the 2D 
metal is not a Fermi liquid.

\vspace{1.cm}
\noindent {\bf Acknowledgements}
\vspace{0.5cm}

The authors are grateful to E. Abrahams, V. Dobrosavljevi\'{c}, and E. Miranda
for useful discussions.  This work was supported by NSF Grant No. DMR-9510355.

\end{document}